\documentstyle[psfig,conf_iap]{article}
\begin{document}
\heading{%
%Begin Heading
%
Confusion noise in millimetre/submillimetre surveys
%
%End Heading
} 
\par\medskip\noindent
\author{%
%Begin Author names
A.\,W. Blain,$^{1}$ R.\,J. Ivison,$^{2}$ Ian Smail,$^{3}$ J.-P. Kneib$^{4}$
%End Author names
}
\address{%
%First address
Cavendish Laboratory, Madingley Road, Cambridge, UK. 
}
\address{%
% Second Address
Institute for Astronomy, University of Edinburgh, Blackford Hill, Edinburgh, UK. 
}
\address{%
% Third Address
Department of Physics, University of Durham, South Road, Durham, UK. 
}
\address{%
% Fourth Address
Observatoire Midi-Pyrenees, 14 Avenue E. Belin, 31400 Toulouse, France. 
}

\begin{abstract}
The flux density from undetected sources in the observing beam of a telescope
produces source confusion noise in the resulting maps of the sky, and thus limits
sensitivity. In a recent paper \cite{BIS} we discussed this effect in the 
millimetre/submillimetre (mm/submm) waveband, using a simple model of galaxy 
evolution that could account for the gravitationally lensed images of distant 
dusty galaxies newly discovered in the fields of clusters of galaxies \cite{SIB}. 
New models explain all the available mm, submm and far-infrared data 
\cite{BSIK}. The associated predictions of source confusion noise are presented 
here. The new and old estimates agree to within a factor of two on angular 
scales between about 0.3 and 50\,arcmin, and are most similar on angular scales 
of about 1\,arcmin. The new estimates are greater/less than the old ones on 
larger/smaller angular scales. 
\end{abstract}

\medskip

Source confusion noise is a significant concern for both galaxy surveys and 
cosmic microwave background radiation (CMBR) anisotropy measurements in the 
mm/submm waveband. The dominant source of confusion noise is expected to be 
distant dusty star-forming galaxies and active galactic nuclei (AGN). The 
population of these objects was determined for the first time last year \cite{SIB}, 
and used to make the first estimates of mm/submm-wave confusion noise 
\cite{BIS} based on direct observations. 

By modeling the results of the observations in more detail, and taking into 
account the recently measured spectrum of extragalactic background radiation 
in the mm, submm and far-infrared wavebands, the form of evolution of distant 
dusty galaxies has now been determined more accurately. Families of 
galaxy evolution models that are consistent with all the observations are 
described elsewhere \cite{BSIK}. These new results allow us to derive more 
definitive estimates of source confusion noise, and to determine the 
model-dependent uncertainties in the estimates: see Fig.\,1. More details and 
references to the range of ground-based and space-borne telescopes and 
instruments included in Fig.\,1 can be found in \cite{BIS}. 

At longer wavelengths in the mm waveband the contribution of non-thermal 
radio sources to the population of confusing sources is expected to dominate 
that of dusty galaxies and AGN. The importance of confusing radio sources at 
frequencies from about 10 to 100\,GHz is rather uncertain at present, but is 
under active investigation because of its significance for CMBR anisotropy 
measurements.

%\vskip -2cm
\begin{figure}[t]
\psfig{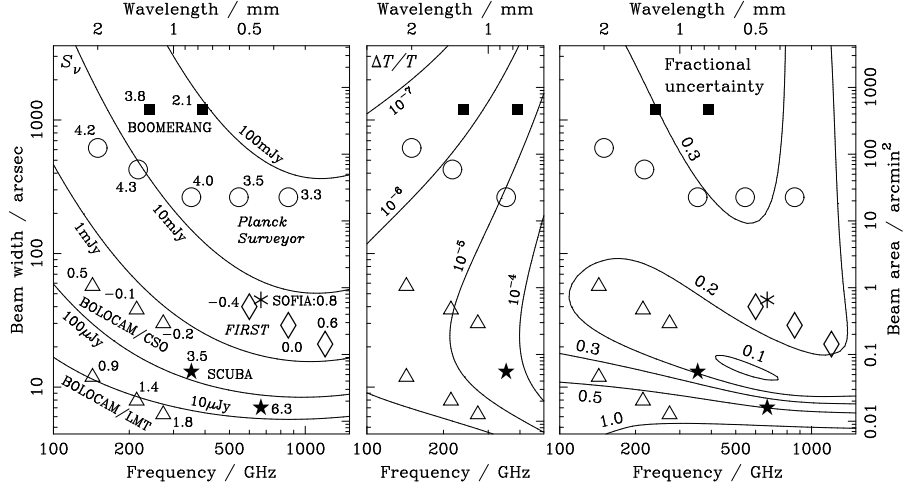}
\vskip -0.25cm
\caption[]{1$\sigma$ confusion noise as a function of both observing frequency 
and angular scale, presented in units of noise equivalent flux density $S_\nu$ 
(left) and relative CMBR temperature uncertainty $\Delta T / T$ (centre). The 
estimated model-dependent uncertainty in the results is shown on the right.  
The resolution limits and observing frequencies of several telescopes are also 
plotted using different point styles. The two-digit numbers alongside the points 
represent the logarithms of the integration times in hours after which confusion 
noise is expected to exceed instrumental noise for each telescope; for 
comparison, 1\,day, 30\,days and 1\,year correspond to values of 1.38, 2.86 and 
3.94 respectively. These times refer to a single-pointed imaging observation, 
except for {\it Planck Surveyor}, for which the times refer to the noise per pixel 
expected in a 14-month all-sky survey. The resolution limit of a large 
interferometer array, such as the MMA, would lie below the bottom of the panels. 
The corresponding two-digit numbers for the MMA are -0.1 at 353\,GHz and 
4.1 at 230\,GHz, assuming a 3-arcsec beam. For details of the instruments 
see \cite{BIS}.}
\end{figure}
%\vskip -2cm
%

This paper provides the latest estimates of source confusion noise due to distant 
dusty galaxies and AGN in the mm, submm and far-infrared wavebands. The 
accuracy of the predictions as a function of the angular resolution and frequency 
of observations is assessed for the first time.

\vfill
\end{document}